\documentstyle[12pt,prd,aps,epsfig,preprint]{revtex}
\begin{document}
\draft
\date{\today}
\title{Scalar $\sigma$ meson effects in radiative  $\rho^0$-meson decays}

\author{A. Gokalp~\thanks{agokalp@metu.edu.tr},  S. Solmaz and O. Yilmaz~\thanks{oyilmaz@metu.edu.tr}}
\address{ {\it Physics Department, Middle East Technical University,
06531 Ankara, Turkey}} \maketitle

\begin{abstract}
We study the radiative $\rho^0\rightarrow \pi^+\pi^-\gamma$ and
$\rho^0\rightarrow \pi^0\pi^0\gamma$ decays and we calculate their
branching ratios using a phenomenological approach by adding to
the amplitude calculated within the framework of chiral
perturbation theory and vector meson dominance the amplitude of
$\sigma$-meson intermediate state. Our results for the branching
ratios are in good agreement with the experimental values.
\end{abstract}

\thispagestyle{empty} ~~~~\\ 
\newpage
\setcounter{page}{1}
The radiative decays of neutral vector mesons into a single photon
and a pair of neutral pseudoscalar mesons have been a subject of
continuous interest. The studies of such decays may serve as tests
for the theoretical ideas about the nature of the intermediate
states and the interesting mechanisms of these decays, and they
may thus provide information about the complicated dynamics of
meson physics in the low energy region.

The very recent measurement of the branching ratio for the decay
$\rho^{0}\rightarrow\pi^{0}\pi^{0}\gamma$ by the SND Collaboration
obtained the value
$BR(\rho^{0}\rightarrow\pi^{0}\pi^{0}\gamma)=(4.1^{+1.0}_{-0.9}\pm
0.3)\times 10^{-5}$  \cite{R1}, thus improving their previous
preliminary report of
$BR(\rho^{0}\rightarrow\pi^{0}\pi^{0}\gamma)=(4.8^{+3.4}_{-1.8}\pm
0.3)\times 10^{-5}$  \cite{R2}. On the other hand, the branching
ratio for the decay $\rho^{0}\rightarrow\pi^{+}\pi^{-}\gamma$ was
reported earlier by Novosibirsk group  as
$BR(\rho^{0}\rightarrow\pi^{+}\pi^{-}\gamma)=(9.9\pm 1.6)\times
10^{-3}$ \cite{R3,R4}, and it was observed that the pion
bremsstrahlung is the main mechanism for this decay with the
structural radiation proceeding through the intermediate scalar
resonance making less than one-order of magnitude smaller
contribution to the branching ratio \cite{R3}.

The theoretical studies of radiative $\rho$-meson decays was
initiated by Singer \cite{R5} who calculated the amplitude for the
decay $\rho^{0}\rightarrow\pi^{+}\pi^{-}\gamma$ by considering
only the bremsstrahlung mechanism, and he assumed that the decay
$\rho^{0}\rightarrow\pi^{0}\pi^{0}\gamma$ proceeds through an
($\omega\pi$) intermediate state as $\rho^{0}\rightarrow
(\omega)\pi^0 \rightarrow(\pi^{0}\gamma)\pi^{0}$. The vector meson
dominance (VMD) calculation of Bramon et al. \cite{R6} with this
intermediate state using standard Lagrangians obeying
SU(3)-symmetry gave the value
B($\rho^{0}\rightarrow\pi^{0}\pi^{0}\gamma$)=$1.1\times 10^{-5}$
for the branching ratio. However, they also noted that final state
interactions could lead to a larger value for the branching ratio
B($\rho^{0}\rightarrow\pi^{0}\pi^{0}\gamma$) through the mechanism
$\rho^{0}\rightarrow(\pi^{+}\pi^{-})\gamma\rightarrow(\pi^{0}\pi^{0})\gamma$.
Bramon et al. \cite{R7} later considered the radiative vector
meson decays within the framework of chiral effective Lagrangians
enlarged to include on-shell vector mesons using chiral
perturbation theory, and they calculated the branching ratios for
various decays at the one-loop level, including both $\pi\pi$ and
$K\bar{K}$ intermediate loops. In this approach, the decay
$\rho^{0}\rightarrow\pi^{0}\pi^{0}\gamma$ proceeds mainly through
the charged pion $(\pi^+\pi^-)$loops, contribution of kaon-loops
being three orders of magnitude smaller, resulting in the decay
rate $\Gamma(\rho^{0}\rightarrow\pi^{0}\pi^{0}\gamma)_\chi$=$1.42$
KeV which is of the same order of magnitude as the VMD
contribution. The interference between the pion-loop contribution
and the VMD-amplitude turns out to be constructive leading to
BR$(\rho^{0}\rightarrow\pi^{0}\pi^{0}\gamma)_{VMD+\chi}$=$2.6\times
10^{-5}$. However, this value is still substantially smaller than
the latest experimental result quoted above.

Since the experimental result for the branching ratio
B$(\rho^{0}\rightarrow\pi^{0}\pi^{0}\gamma$) is almost nearly
twice the theoretical value calculated using VMD and chiral-loop
amplitudes, the mechanism of the decay
$\rho^{0}\rightarrow\pi^{0}\pi^{0}\gamma$ should reexamined and
additional contributions should be investigated. One additional
contribution to the decay may be provided by the amplitude
involving scalar-isoscalar $\sigma$-meson as an intermediate
state. Although the existence of $\sigma$-meson has long been
controversial, an increasing number of theoretical and
experimental analyzes find a $\sigma$-pole position near
$(500-i250)$ MeV \cite{R8}. Furthermore, the
$D^+\rightarrow\sigma\pi^0\rightarrow 3\pi$ decay channel observed
by the Fermilab (E791) Collaboration is interpreted to provide
direct experimental evidence for $\sigma$-meson where it is seen
as a clear dominant peak with $M_{\sigma}=(478^{+24}_{-23}\pm 17)$
MeV, and $\Gamma_\sigma=(324^{+42}_{-40}\pm 21)$ MeV \cite{R9}.
Since $\sigma$-meson is assumed  to couple strongly to low mass
pion pairs, the $\rho^{0}\rightarrow\pi\pi\gamma$ decays thus
provide an opportunity to investigate  the theoretical ideas about
the role $\sigma$-meson plays in the dynamics of low energy meson
physics.

One way to include the effect of $\sigma$-meson in the decay
mechanisms of $\rho^{0}\rightarrow\pi^{+}\pi^{-}\gamma$ and
$\rho^{0}\rightarrow\pi^{0}\pi^{0}\gamma$ decays is to consider
its contribution as resulting from a $\sigma$-pole intermediate
state, that is to assume that the contributions of $\sigma$-meson
to the decay mechanisms of these decays result from the
corresponding amplitudes of
$\rho^{0}\rightarrow(\sigma\gamma)\rightarrow(\pi^{+}\pi^{-})\gamma$
and
$\rho^{0}\rightarrow(\sigma\gamma)\rightarrow(\pi^{0}\pi^{0})\gamma$
reactions. In a previous work \cite{R10}, two of the present
authors calculated the branching ratio
B$(\rho^{0}\rightarrow\pi^{+}\pi^{-}\gamma$)in a phenomenological
framework using pion bremsstrahlung amplitude and $\sigma$-meson
pole amplitude. The experimental value of this branching ratio was
then used to calculate the coupling constant
$g_{\rho\sigma\gamma}$ as a function of $\sigma$-meson parameters
$M_{\sigma}$ and $\Gamma_\sigma$. These authors in a following
work \cite{R11} calculated the branching ratio
B$(\rho^{0}\rightarrow\pi^{0}\pi^{0}\gamma$) using the values of
the coupling constant $g_{\rho\sigma\gamma}$ thus obtained in a
phenomenological approach where the contributions of
$\sigma$-meson, $w$-meson intermediate states and of the
pion-loops are considered. The branching ratio
B$(\rho^{0}\rightarrow\pi^{0}\pi^{0}\gamma$) obtained this way for
$M_{\sigma}=478$ MeV and $\Gamma_\sigma=324$ MeV was more than an
order of magnitude larger than the experimental value. This
unrealistic value was the result of the constant
$\rho\rightarrow\sigma\gamma$ amplitude used and consequently the
large coupling constant $g_{\rho\sigma\gamma}$ extracted from the
experimental value of the branching ratio of the
$\rho^{0}\rightarrow\pi^{+}\pi^{-}\gamma$ decay. Therefore, it may
be concluded that it is not realistic to include the
$\sigma$-meson in the mechanisms of the radiative $\rho^0$-meson
decays as an intermediate pole state.

On the other hand, $\rho^{0}\rightarrow\pi^{0}\pi^{0}\gamma$ decay
was also considered by Marco et al. \cite{R12} in the framework of
unitarized chiral perturbation theory. They noted that the
energies of two pion system are quite large so that the decay
cannot be treated with standard chiral perturbation theory. They
used the techniques of chiral unitary theory developed earlier, a
review is given  by Oller et al. \cite{R13}, to include the final
state interactions of two pions by summing the pion-loops through
the Bethe-Salpeter equation. The branching ratio for
$\rho^{0}\rightarrow\pi^{0}\pi^{0}\gamma$ they obtained was
B$(\rho^{0}\rightarrow\pi^{0}\pi^{0}\gamma$)=$1.4\times 10^{-5}$
and they furthermore noted that this result could be interpreted
as the result of the mechanism
$\rho^{0}\rightarrow(\sigma)\gamma\rightarrow
(\pi^{0}\pi^{0})\gamma$ since $\pi^0\pi^0$ interaction is
dominated by $\sigma$-pole in the relevant energy regime of this
decay.

Thus, it seems that a natural way to include the effects of
$\sigma$-meson in the mechanisms of radiative $\rho^0$-meson
decays is to assume that the $\sigma$-meson couples to
$\rho^0$-meson through the pion-loop. In this work, we reconsider
the approach used in the references \cite{R10} and \cite{R11}, and
we study the contribution of the $\sigma$-meson intermediate state
amplitude to $\rho^{0}\rightarrow\pi^{+}\pi^{-}\gamma$ and
$\rho^{0}\rightarrow\pi^{0}\pi^{0}\gamma$ decays. We follow a
phenomenological approach and assume that the $\sigma$-meson
couples to the $\rho^0$-meson through a pion-loop, in other words
we assume that the amplitude $\rho^0\rightarrow\sigma\gamma$
results from the sequential
$\rho^{0}\rightarrow(\pi^{+}\pi^{-})\gamma\rightarrow\sigma\gamma$
mechanism  as suggested by the unitarized  chiral perturbation
theory in which $\sigma$-meson is generated dynamically by
unitarizing the one-loop pion amplitudes. We use the coupling
constants that are determined from the experimental values of the
relevant quantities that we calculate employing the effective
Lagrangians of our approach. Although the decay
$\rho^{0}\rightarrow\pi^{+}\pi^{-}\gamma$ is dominated by the
pion-bremsstrahlung amplitude, for the consistency of our
formalism we study both of
$\rho^{0}\rightarrow\pi^{+}\pi^{-}\gamma$ and
$\rho^{0}\rightarrow\pi^{0}\pi^{0}\gamma$ decays and calculate
their branching ratios.

Our calculation is based on the Feynman diagrams shown in Fig. 1
for $\rho^{0}\rightarrow\pi^{+}\pi^{-}\gamma$ decay and in Fig. 2
for $\rho^{0}\rightarrow\pi^{0}\pi^{0}\gamma$ decay. The last
diagrams in Fig. 1 a, b, c and in Fig. 2 b, c are the direct terms
required to establish the gauge invariance. We describe the
$\omega\rho\pi$-vertex by the effective Lagrangian

\begin{eqnarray} \label{e1}
{\cal L}^{eff}_{\rho\omega\pi}=g_{\omega\rho\pi}
\epsilon^{\mu\nu\alpha\beta}\partial_{\mu}\omega_{\nu}
\partial_{\alpha}\vec{\rho}_{\beta}\cdot\vec{\pi}~~,
\end{eqnarray}
which also conventionally defines the coupling constant
$g_{\omega\rho\pi}$. Achasov et al. \cite{R14} assumed that
$\omega\rightarrow 3\pi$ decay proceeds with the intermediate
$\rho\pi$ state as
$\omega\rightarrow(\rho)\pi\rightarrow\pi\pi\pi$ and using
experimental value of the $\omega\rightarrow 3\pi$ width they
determined this coupling constant as $g_{\rho\omega\pi}=(14.3\pm
0.2)~~~GeV^{-1}$. The $\rho\pi\pi$-vertex is described by the
effective Lagrangian
\begin{eqnarray}\label{e2}
{\cal L}^{eff}_{\rho\pi\pi}=g_{\rho\pi\pi}
\vec{\rho}_{\mu}\cdot(\partial^{\mu}\vec{\pi}\times\vec{\pi})~~.
\end{eqnarray}
The experimental decay width of the decay $\rho\rightarrow\pi\pi$
\cite{R4} then yields the value $g_{\rho\pi\pi}=(6.03\pm 0.02)$
for the coupling constant $g_{\rho\pi\pi}$.  For the
$\sigma\pi\pi$-vertex we use the effective Lagrangian
\begin{eqnarray} \label{e3}
{\cal L}^{eff}_{\sigma\pi\pi}=
\frac{1}{2}g_{\sigma\pi\pi}M_{\sigma}\vec{\pi}\cdot\vec{\pi}\sigma~~.
\end{eqnarray}
Using the experimental values for $M_{\sigma}$ and $\Gamma_\sigma$
\cite{R9} as $M_{\sigma}=(483\pm 31)$ MeV and
$\Gamma_\sigma=(338\pm 48)$ MeV, where statistical and systematic
errors are added in quadrature \cite{R15}, we obtain the strong
coupling constant $g_{\sigma\pi\pi}$ as $g_{\sigma\pi\pi}=(5.34\pm
0.55)$. We note that our effective Lagrangians ${\cal
L}^{eff}_{\sigma\pi\pi}$ and ${\cal L}^{eff}_{\rho\pi\pi}$ are the
ones that result from an extension of the $\sigma$-model to
include the isovector $\rho$ through a Yang-Mills local gauge
theory based on isospin with the vector meson mass generated
through the Higgs mechanism \cite{R16}. The
$\omega\pi\gamma$-vertex is described by the effective Lagrangian
\begin{eqnarray}\label{e5}
{\cal L}^{eff}_{\omega\pi\gamma}=g_{\omega\pi\gamma}
\epsilon^{\mu\nu\alpha\beta}\partial_{\mu}\omega_{\nu}
\partial_{\alpha}A_{\beta}\pi~~.
\end{eqnarray}
We then obtain the coupling constant $g_{\omega\pi\gamma}$ from
the experimental partial width \cite{R4} of the radiative decay
$\omega\rightarrow\pi^{0}\gamma$ as $g_{\omega\pi\gamma}=(0.706\pm
0.021)~~GeV^{-1}$.

Meson-meson interactions were studied by Oller and Oset \cite{R17}
using the standard Chiral Lagrangian in lowest order of chiral
perturbation theory that contains the most general low energy
interactions of the pseudoscalar meson octet in this order. We use
their results for the four pseudoscalar amplitudes
$\pi^+\pi^-\rightarrow\pi^0\pi^0$ and
$\pi^+\pi^-\rightarrow\pi^+\pi^-$ that we need in the loop
diagrams in Fig. 1 b and in Fig. 2 b. We note that as shown by
Oller \cite{R18} due to gauge invariance the off-shell parts of
the amplitudes, that should be kept inside the loop integration,
do not contribute, and as a result the amplitudes ${\cal
M}_{\chi}(\pi^+\pi^-\rightarrow\pi^0\pi^0)$ and ${\cal
M}_{\chi}(\pi^+\pi^-\rightarrow\pi^+\pi^-)$ factorize in the
expressions for the loop diagrams.

In our calculation of the invariant amplitude, in the
$\sigma$-meson propagator we make the replacement
$q^2-M^2\rightarrow q^2-M^2+iM\Gamma$ and we use the energy
dependent width for $\sigma$-meson which is given as
\begin{eqnarray}\label{e6}
\Gamma_\sigma (q^2)=\Gamma_\sigma
\frac{M_\sigma^2}{q^2}\sqrt{\frac{q^2-4M_\pi^2}{M_\sigma^2-4M_\pi^2}}
\theta(q^2-4M_\pi^2)~~.
\end{eqnarray}

For the loop integrals appearing in Figs. 1 and 2 we use the
results of Lucio and Pestiau \cite{R19} who evaluated similar
integrals using dimensional regularization. In our case the
contribution of the pion-loop amplitude corresponding to
$\rho^0\rightarrow(\pi^+\pi^-)\gamma\rightarrow\pi^0\pi^0\gamma$
reaction can be written as
\begin{eqnarray}\label{e7}
{\cal M_\pi}=-\frac{e~g_{\rho\pi\pi}{\cal
M}_{\chi}(\pi^+\pi^-\rightarrow\pi^0\pi^0)}{2\pi^{2}M_{\pi}^{2}}I(a,b)
\left[(p\cdot k)(\epsilon\cdot u)-(p\cdot\epsilon)(k\cdot
u)\right]~~,
\end{eqnarray}
where $a=M_\rho^2/M_\pi^2$,~ $b=(p-k)^2/M_\pi^{2}$,~ ${\cal
M}_{\chi}(\pi^+\pi^-\rightarrow\pi^0\pi^0)=-(s-M_\pi^2)/f_\pi^2$,~
$s=M_{\pi^0\pi^0}^2$,~ $f_\pi=92.4$ MeV, $p(u)$ and $k(\epsilon)$
being the momentum (polarization vector) of $\rho$-meson and
photon, respectively. The amplitudes corresponding to
$\rho^0\rightarrow(\pi^+\pi^-)\gamma\rightarrow\pi^+\pi^-\gamma$,
$\rho^0\rightarrow(\pi^+\pi^-)\gamma\sigma\rightarrow\gamma\pi^+\pi^-$
and
$\rho^0\rightarrow(\pi^+\pi^-)\gamma\sigma\rightarrow\gamma\pi^0\pi^0$
reactions can similarly be written. The function I(a,b) is given
as
\begin{eqnarray}\label{e8}
I(a,b)=\frac{1}{2(a-b)} -\frac{2}{(a-b)^{2}}\left [
f\left(\frac{1}{b}\right)-f\left(\frac{1}{a}\right)\right ]
+\frac{a}{(a-b)^{2}}\left [
g\left(\frac{1}{b}\right)-g\left(\frac{1}{a}\right)\right ]
\end{eqnarray}
where
\begin{eqnarray}\label{e9}
&&f(x)=\left \{
\begin{array}{rr}
           -\left [ \arcsin (\frac{1}{2\sqrt{x}})\right ]^{2}~,& ~~x>\frac{1}{4} \\
\frac{1}{4}\left [ \ln (\frac{\eta_{+}}{\eta_{-}})-i\pi\right
]^{2}~, & ~~x<\frac{1}{4}
            \end{array} \right.
\nonumber \\ && \nonumber \\ &&g(x)=\left \{ \begin{array}{rr}
        (4x-1)^{\frac{1}{2}} \arcsin(\frac{1}{2\sqrt{x}})~, & ~~ x>\frac{1}{4} \\
 \frac{1}{2}(1-4x)^{\frac{1}{2}}\left [\ln (\frac{\eta_{+}}{\eta_{-}})-i\pi \right ]~, & ~~ x<\frac{1}{4}
            \end{array} \right.
\nonumber \\ && \nonumber \\ &&\eta_{\pm}=\frac{1}{2x}\left [
1\pm(1-4x)^{\frac{1}{2}}\right ] ~.
\end{eqnarray}
We then calculate the invariant amplitude ${\cal M}$(E$_{\gamma}$,
E$_{1}$) from the corresponding Feynman diagrams shown in Fig. 1
and 2 for the decays $\rho^{0}\rightarrow\pi^{+}\pi^{-}\gamma$ and
$\rho^{0}\rightarrow\pi^{0}\pi^{0}\gamma$, respectively.The
differential decay probability for an unpolarized $\rho^{0}$-meson
at rest is then given as
\begin{eqnarray}\label{e10}
\frac{d\Gamma}{dE_{\gamma}dE_{1}}=\frac{1}{(2\pi)^{3}}~\frac{1}{8M_{\rho}}~
\mid {\cal M}\mid^{2} ,
\end{eqnarray}
where E$_{\gamma}$ and E$_{1}$ are the photon and pion energies
respectively. We perform an average over the spin states of
$\rho^{0}$-meson and a sum over the polarization states of the
photon. The decay width is then obtained by integration
\begin{eqnarray}\label{e11}
\Gamma=\left(\frac{1}{2}\right)\int_{E_{\gamma,min.}}^{E_{\gamma,max.}}dE_{\gamma}
       \int_{E_{1,min.}}^{E_{1,max.}}dE_{1}\frac{d\Gamma}{dE_{\gamma}dE_{1}}
\end{eqnarray}
where the factor ($\frac{1}{2}$) is included for the calculation
of the decay rate for $\rho^{0}\rightarrow\pi^{0}\pi^{0}\gamma$
because of the $\pi^{0}\pi^{0}$ in the final state. The minimum
photon energy is E$_{\gamma, min.}=0$ and the maximum photon
energy is given as
$E_{\gamma,max.}=(M_{\rho}^{2}-4M_{\pi}^{2})/2M_{\rho}$=338 MeV.
The maximum and minimum values for pion energy E$_{1}$ are given
by
\begin{eqnarray}\label{e12}
\frac{1}{2(2E_{\gamma}M_{\rho}-M_{\rho}^{2})}
[ -2E_{\gamma}^{2}M_{\rho}+3E_{\gamma}M_{\rho}^{2}-M_{\rho}^{3}
 ~~~~~~~~~~~~~~~~~~~~~~~~~~~~ \nonumber \\
\pm  E_{\gamma}\sqrt{(-2E_{\gamma}M_{\rho}+M_{\rho}^{2})
       (-2E_{\gamma}M_{\rho}+M_{\rho}^{2}-4M_{\pi}^{2})}~] ~.
\nonumber
\end{eqnarray}

The photon spectra for the branching ratio of the decay
$\rho^{0}\rightarrow \pi^+\pi^-\gamma$ are plotted in Fig. 3 as a
function of photon energy $E_\gamma$. The contributions of
bremsstrahlung and structural radiation amplitudes calculated with
pion-loop and with $\sigma$-meson intermediate state as well as
the contribution of the interference term are shown as a function
of the photon energy, where the minimum photon energy is taken as
$E_{\gamma, min}=50$ MeV since the experimental value of the
branching ratio is determined for this range of photon energies
\cite{R3}. On the same figure we also show the experimental data
points taken from Ref.\cite{R3} which are normalized to our
results. As it can be seen in Fig. 3, the shape of the photon
energy distribution is reproduced well. As expected, the main
contribution to the branching ratio comes from the
pion-bremsstrahlung amplitude, the contributions of pion-loop and
$\sigma$-meson intermediate states becoming noticeable only in the
region of high photon energies. On the other hand, if a
$\sigma$-meson pole model or equivalently a constant
$\rho\rightarrow\sigma\gamma$ amplitude is used as in Ref.
\cite{R10} the contribution of the $\sigma$-term becomes
increasingly important in the region of high photon energies
dominating the contribution of the bremsstrahlung amplitude, and
although its contribution is somewhat reduced by the interference
term the $\sigma$-meson amplitude makes the main contribution to
the branching ratio in the region of high photon energies
conflicting with the experimental spectrum. For the contribution
of different amplitudes to the branching ratio, we obtain
B$(\rho^{0}\rightarrow \pi^+\pi^-\gamma)_\gamma=(1.14\pm
0.01)\times 10^{-2}$ from the bremsstrahlung amplitude,
B$(\rho^{0}\rightarrow \pi^+\pi^-\gamma)_\pi=(0.45\pm 0.08)\times
10^{-5}$ from the pion-loop amplitude and B$(\rho^{0}\rightarrow
\pi^+\pi^-\gamma)_\sigma=(0.83\pm 0.16)\times 10^{-4}$ from the
$\sigma$-meson intermediate state. If we consider also the
interference between the $\pi$-loop and $\sigma$-meson amplitudes,
we then obtain the contribution coming from the structural
radiation as B$(\rho^{0}\rightarrow \pi^+\pi^-\gamma)=(0.83\pm
0.14)\times 10^{-4}$  which is reasonable  agreement with the
experimental limit B$(\rho^{0}\rightarrow
\pi^+\pi^-\gamma)<5\times 10^{-3}$ deduced by Dolinsky et al.
\cite{R3} for the structural radiation. Dolinsky et al. \cite{R3}
also extracted the experimental limit
B$(\rho^{0}\rightarrow\epsilon(700)\gamma\rightarrow
\pi^+\pi^-\gamma)<4\times 10^{-4}$  where the transition proceeds
through the intermediate scalar resonance which is considerably
lower. Our result for the contribution of the $\sigma$-meson
intermediate state B$(\rho^{0}\rightarrow
\pi^+\pi^-\gamma)_\sigma=(0.83\pm 0.16)\times 10^{-4}$ is also in
good agreement with this experimental limit. For the total
contribution, we obtain the branching ratio of the decay
$\rho^{0}\rightarrow \pi^+\pi^-\gamma$  as B$(\rho^{0}\rightarrow
\pi^+\pi^-\gamma)=(1.22\pm 0.02)\times 10^{-2}$ for $E_\gamma>50$
MeV. Our result is in reasonably good agreement with the
experimental number B$(\rho^{0}\rightarrow
\pi^+\pi^-\gamma)=(0.99\pm 0.16)\times 10^{-2}$ \cite{R3}.

The photon spectra resulting from our calculation for the
branching ratio of the $\rho^{0}\rightarrow \pi^0\pi^0\gamma$
decay is shown in Fig. 4. The contributions of VMD-amplitude,
pion-loop amplitude and $\sigma$-meson intermediate state
amplitude as well as the total interference term are indicated.
The contributions of VMD amplitude, pion-loop amplitude and
$\sigma$-meson intermediate state amplitude to the branching ratio
of the decay are
B$(\rho^{0}\rightarrow\pi^{0}\pi^{0}\gamma)_{VMD}=(1.03\pm
0.02)\times 10^{-5}$, B$(\rho^{0}\rightarrow
\pi^{0}\pi^{0}\gamma)_\pi=(1.07\pm 0.02)\times 10^{-5}$ and
B$(\rho^{0}\rightarrow\pi^{0}\pi^{0}\gamma)_\sigma=(4.96\pm
0.18)\times 10^{-5}$ respectively. We see that $\sigma$-meson
intermediate state makes an important contribution to the
branching ratio comparable to the contributions of VMD- and
pion-loop amplitudes. The values we obtain for
B$(\rho^{0}\rightarrow \pi^{0}\pi^{0}\gamma)_{VMD}$ and for
B$(\rho^{0}\rightarrow \pi^{0}\pi^{0}\gamma)_{\pi}$ are in
agreement with previous calculations \cite{R6,R7}. For the total
branching ratio, including interference terms, we obtain the
result B$(\rho^{0}\rightarrow\rightarrow
\pi^{0}\pi^{0}\gamma)=(4.95\pm 0.82)\times 10^{-5}$ which is in a
better agreement  with the experimental result
B$(\rho^{0}\rightarrow \pi^{0}\pi^{0}\gamma)=(4.1^{+1.0}_{-0.9}\pm
0.3)\times 10^{-5}$ \cite{R1} than the theoretical value obtained
by using only VMD- and chiral pion-loop amplitudes \cite{R7}.

In our work, we follow a phenomenological approach and in a model
for the decay mechanism of $\rho^0$-meson radiative decays
including the contribution coming from the $\sigma$-meson
intermediate state as well as VMD- and chiral pion-loop
contributions we calculate the branching ratios for the
$\rho^{0}\rightarrow \pi^+\pi^-\gamma$ and $\rho^{0}\rightarrow
\pi^0\pi^0\gamma$ decays. In our calculations of the branching
ratios, the coupling constants that we use in our model are
determined from the relevant experimental quantities. Our results
for the branching ratios are in good agreement with the
experimental values, and we believe that our study demonstrates
that the contribution coming from the $\sigma$-meson intermediate
state amplitude should be included in the analysis of radiative
$\rho^0$-meson decays, and moreover $\sigma$-meson should be
considered to couple to the $\rho^0$ meson through a pion-loop.

In a recent paper, Palomar et al. \cite{R20} also evaluated the
branching ratios of the radiative $\rho^0$ and $\omega$ decays
into $\pi^0\pi^0$ and $\pi^0\eta$. They used the sequential vector
decay mechanisms in addition to chiral loops and $\rho-\omega$
mixing. For the $\rho^{0}\rightarrow \pi^0\pi^0\gamma$ decay
$\rho-\omega$ mixing was negligible, but the branching ratio
obtained with the sum of the sequential and loop mechanisms was
about three times larger than with either mechanism alone, leading
to a result $B(\rho^{0}\rightarrow \pi^0\pi^0\gamma)=4.2\times
10^{-5}$ comparable with the present experimental value.

Therefore, in order to understand the mechanism of the
$\rho^{0}\rightarrow \pi^0\pi^0\gamma$ decay and to obtain insight
into the nature and the properties of the $\sigma$-meson, and the
role it plays in the dynamics of low energy meson physics, further
experimental tests such as the measurements of invariant mass
distributions will be very valuable.

\newpage
\begin{figure}
\vspace*{1.0cm}\hspace{2.0cm}
\epsfig{figure=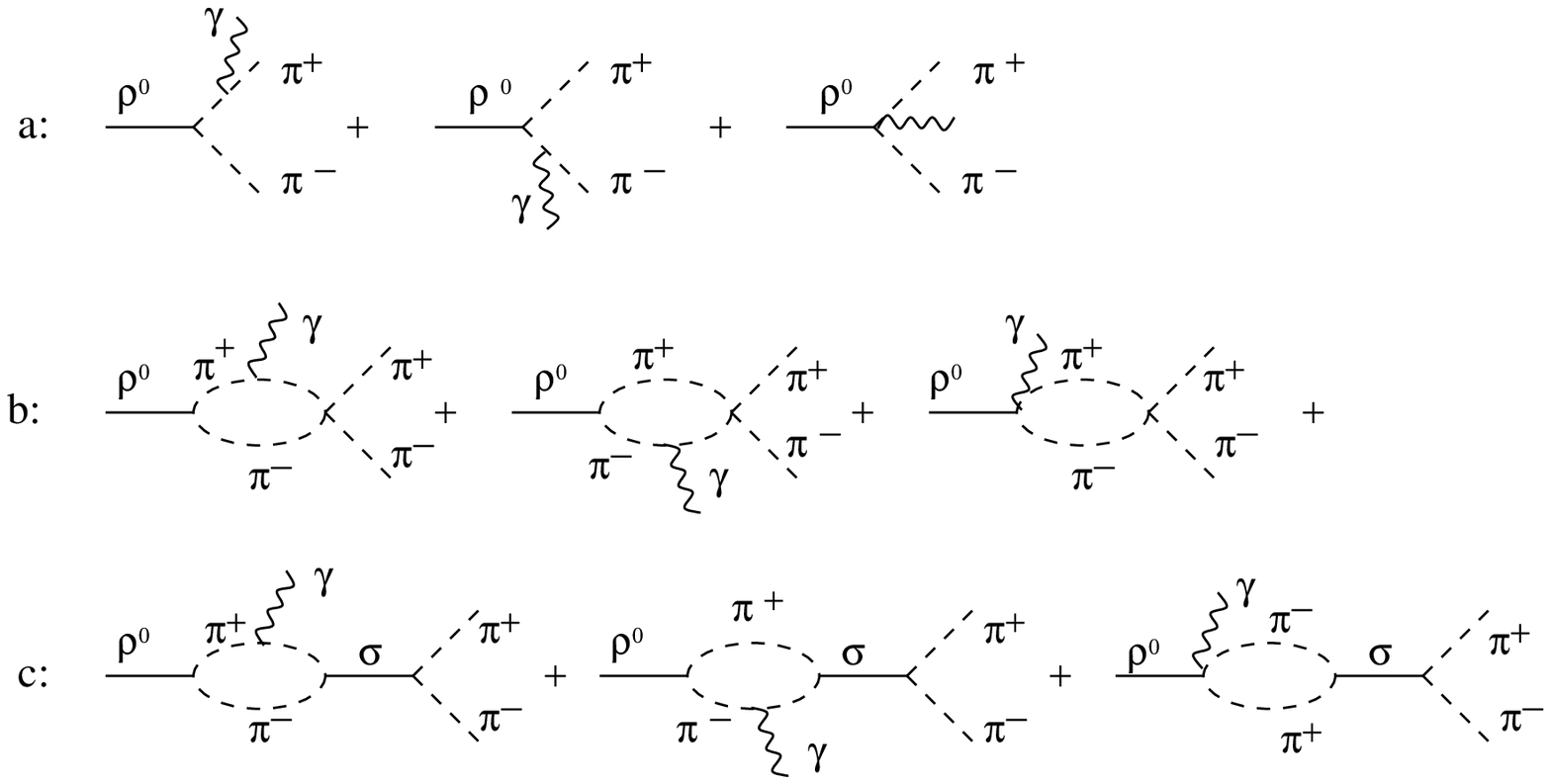,width=12cm,height=7cm,
angle=0}\vspace*{0.0cm} \caption{Feynman diagrams for the decay
$\rho^0\rightarrow\pi^+\pi^-\gamma$.} \label{fig1}
\end{figure}

\begin{figure}\vspace*{1.0cm}\hspace{2.0cm}
\vspace*{1.0cm} \epsfig{figure=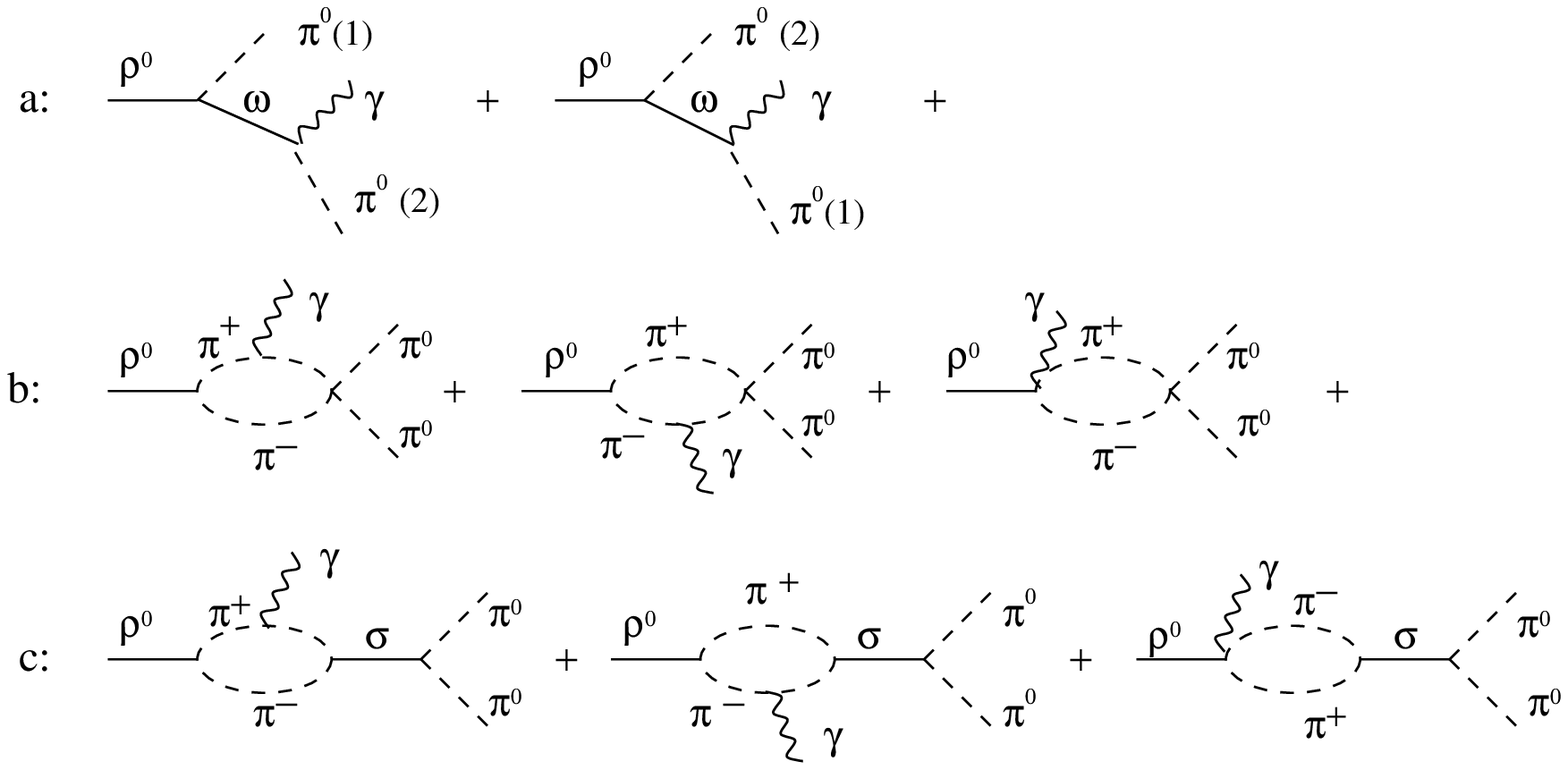,width=12cm,height=7cm,
angle=0}\vspace*{-1.0cm} \caption{Feynman diagrams for the decay
$\rho^0\rightarrow\pi^0\pi^0\gamma$.} \label{fig2}
\end{figure}

\newpage
\begin{figure}
\epsfig{figure=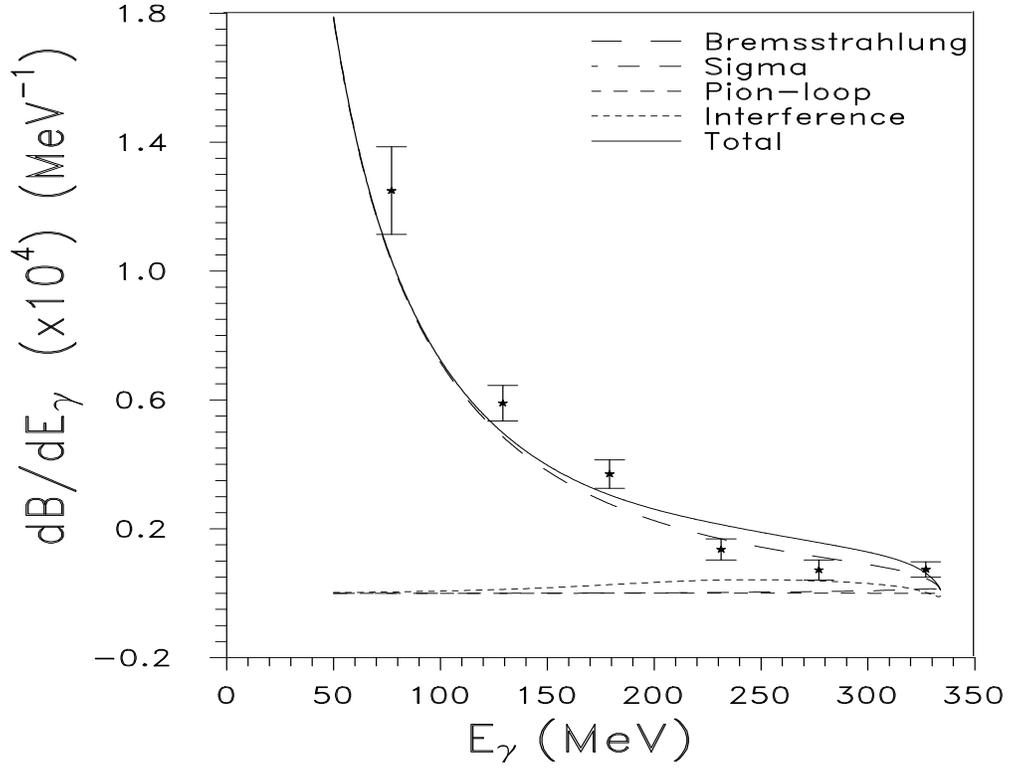,width=15cm,height=15cm}\vspace*{-2.0cm}
\caption{The photon spectra for the branching ratio of
$\rho^0\rightarrow\pi^+\pi^-\gamma$ decay. The contributions of
different terms are indicated. The experimental data taken from
Ref. [3] are normalized to our results.} \label{fig3}
\end{figure}

\begin{figure}
\epsfig{figure=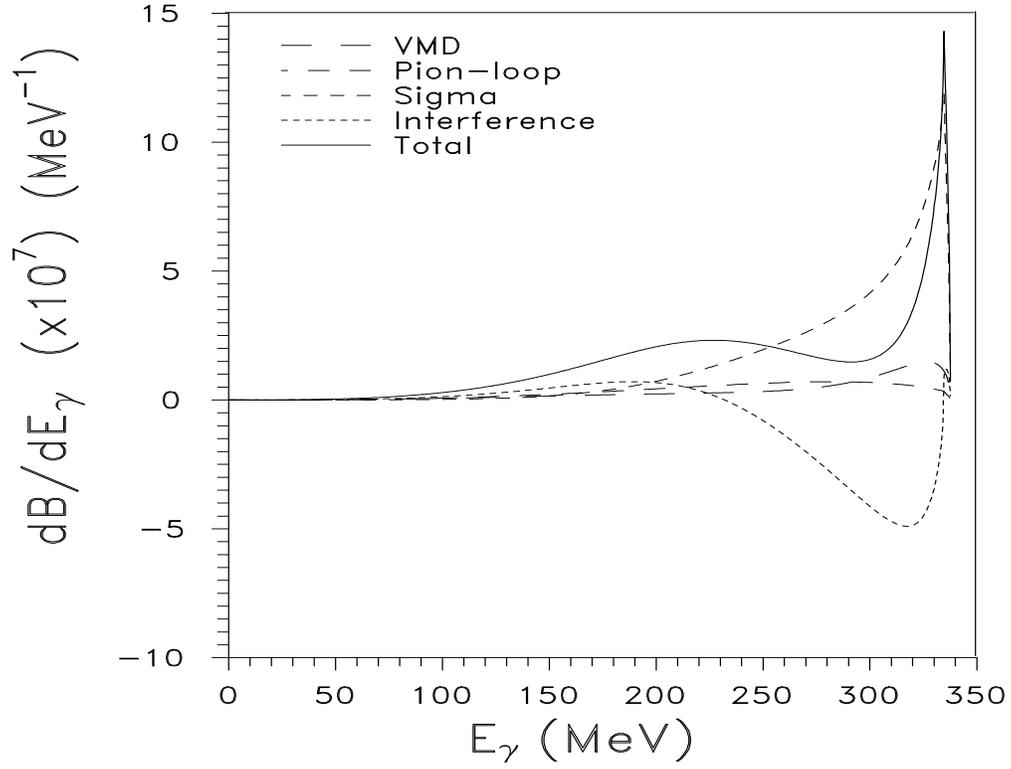,width=15cm,height=15cm} \vspace*{-2.0cm}
\caption{The photon spectra for the branching ratio of
$\rho^0\rightarrow\pi^0\pi^0\gamma$ decay. The contributions of
different terms are indicated. } \label{fig4}
\end{figure}

\end{document}